\documentclass[reprint, superscriptaddress, secnumarabic, amssymb, nobibnotes, aps, prl]{revtex4-2}

\usepackage{graphicx}
\usepackage{epstopdf}
\usepackage[T1]{fontenc}
\usepackage{amsbsy}
\usepackage[utf8]{inputenc}
\usepackage{gensymb}
\usepackage{braket}
\usepackage[normalem]{ulem}
\makeatletter
\def\NAT@biblabelnum#1{#1.}
\makeatother
\makeatletter
\def\fnum@figure{\textbf{Fig}}
\makeatother

\usepackage[T1]{fontenc}
\usepackage{amsmath}
\usepackage{amssymb}
\usepackage{bbm}
\usepackage{braket}
\usepackage{xcolor}
\allowdisplaybreaks
\usepackage{graphicx}
\usepackage[colorlinks=true]{hyperref}  
\hypersetup{
    bookmarks=true,         
    unicode=false,          
    pdftoolbar=true,        
    pdfmenubar=true,        
    pdffitwindow=false,     
    pdfstartview={FitH},    
    pdftitle={titel of paper},    
    pdfauthor={a},     
    pdfsubject={},   
    pdfcreator={},   
    pdfproducer={}, 
    pdfkeywords={} {} {}, 
    pdfnewwindow=true,      
    colorlinks=true,       
    linkcolor=blue, 
    citecolor=blue,        
    filecolor=magenta,      
    urlcolor=blue           
} 
\usepackage[normalem]{ulem}

\begin{document}

\preprint{APS/123-QED}

\title{Ferroelastic exciton splitting in hybrid perovskite nanowalls}

\author{Afreen}
\affiliation{Department of Physics, Indian Institute of Science Education and Research, Bhopal, 462066, India}
\author{J. Delgado-Alvarez}
\affiliation{Nanotechnology on Surfaces $\&$ Plasma Laboratory, Institute of Materials Science of Seville (CSIC-US), C/ Am\'erico
 Vespucio 49, 41092, Seville, Spain}
\author{H. Krishna Mishra}
\affiliation{Nanotechnology on Surfaces $\&$ Plasma Laboratory, Institute of Materials Science of Seville (CSIC-US), C/ Am\'erico
 Vespucio 49, 41092, Seville, Spain}
\author{J. Castillo-Seoane}
\affiliation{Nanotechnology on Surfaces $\&$ Plasma Laboratory, Institute of Materials Science of Seville (CSIC-US), C/ Am\'erico
 Vespucio 49, 41092, Seville, Spain}
\author{Koustav Maiti}
\affiliation{Department of Condensed Matter and Materials Physics, S.N. Bose National Centre for Basic Sciences, Kolkata 700106, India}
\author{Pravrati Taank}
\affiliation{Department of Physics, Indian Institute of Science Education and Research, Bhopal, 462066, India}
\author{A. Borras}
\affiliation{Nanotechnology on Surfaces $\&$ Plasma Laboratory, Institute of Materials Science of Seville (CSIC-US), C/ Am\'erico
 Vespucio 49, 41092, Seville, Spain}
\author{A. Barranco}
\affiliation{Nanotechnology on Surfaces $\&$ Plasma Laboratory, Institute of Materials Science of Seville (CSIC-US), C/ Am\'erico
 Vespucio 49, 41092, Seville, Spain}
\author{Surajit Saha}
\affiliation{Department of Physics, Indian Institute of Science Education and Research, Bhopal, 462066, India}
\author{Priya Mahadevan}
\affiliation{Department of Condensed Matter and Materials Physics, S.N. Bose National Centre for Basic Sciences, Kolkata 700106, India}
\author{J. R. Sanchez-Valencia$^*$}%
\affiliation{Nanotechnology on Surfaces $\&$ Plasma Laboratory, Institute of Materials Science of Seville (CSIC-US), C/ Am\'erico
 Vespucio 49, 41092, Seville, Spain}
\author{K. V. Adarsh$^*$}%
\affiliation{Department of Physics, Indian Institute of Science Education and Research, Bhopal, 462066, India}


\begin{abstract}
\begin{center}

\noindent\text{$^*$Corresponding authors E-mails: \textcolor{blue}{adarsh@iiserb.ac.in} and \textcolor{blue}{ jrsanchez@icmse.csic.es}}
\\
\end{center}
Hybrid metal-halide perovskites are soft semiconductors in which electronic excitations are strongly influenced by lattice distortions and structural phase transitions. An important open question is whether ferroelastic symmetry breaking merely broadens optical resonances or instead modifies excitonic states through exciton-lattice coupling. Here, we address this question using highly aligned MAPbI$_3$ nanowalls fabricated by glancing-angle deposition, enabling symmetry-selective coupling between ferroelastic texture, structural anisotropy, and a well-defined optical axis. Combining temperature-dependent photoluminescence, X-ray diffraction and polarization-resolved ultrafast transient absorption spectroscopy, we observe a polarization-selective excitonic splitting in the orthorhombic phase at 5 K, characterized by orthogonal optical selection rules and a 45 meV energy separation. Near 160 K, where orthorhombic and tetragonal phases coexist, a lower-energy lattice-coupled excitation emerges 58 meV below the centre of the anisotropically split excitonic structure, consistent with coupling between excitonic and lattice-dressed states. At higher temperatures, these excitations progressively acquire lattice-dressed character accompanied by reduced optical anisotropy. A symmetry-guided effective Hamiltonian captures the evolution from anisotropically split excitons to coupled excitonic and lattice-dressed states across the structural transition. Our results show that ferroelastic texture and phase coexistence can modify exciton-lattice coupling, providing a route to symmetry-selective optical responses in soft polar semiconductors.
\end{abstract}

\maketitle


Hybrid metal-halide perovskites occupy an unusual regime of semiconductor physics in which electronic excitations evolve within a structurally soft and dynamically fluctuating lattice environment\cite{li2017chemically,parola2016optical}. Strong spin-orbit coupling, molecular rotational dynamics and low-energy lattice distortions together produce unusually strong carrier-phonon interactions, enabling phenomena ranging from large polarons\cite{miyata2017large,bretschneider2018quantifying,yaffe2017local,park2018excited} to cooperative many-body emission processes such as superfluorescence\cite{biliroglu2022room,findik2021high,poonia2024superfluorescence,aggarwal2025room}. Unlike conventional inorganic semiconductors such as silicon or gallium arsenide\cite{singh2007semiconductor}, the electronic structure of hybrid perovskites remains strongly coupled to the symmetry and dynamics of the underlying lattice. This interplay between structural fluctuations and electronic excitations underlies their unusual optoelectronic properties\cite{peng2025halide,kojima2009organometal,green2014emergence,kong2024efficient,hu2025lead,huang2017lead,buin2014materials,walsh2015self,jasti2024experimental} and makes these materials a compelling platform for investigating how lattice symmetry and structural phase transitions influence excited-state quasiparticles. Yet it remains unclear whether ferroelastic symmetry breaking primarily acts as a source of inhomogeneous broadening or instead modifies excitonic states through exciton-lattice coupling.

A central unresolved question concerns the role of ferroelastic symmetry breaking in shaping excited states in soft semiconductors. The tetragonal to orthorhombic transition in MAPbI$
_3$ produces ferroelastic twin textures, internal strain fields and anisotropic lattice distortions that locally lower crystal symmetry\cite{liu2021ferroic,rohm2019ferroelectric,liu2018chemical,rakita2017tetragonal,kim2015ferroelectric,brivio2015lattice,bari2021ferroelastic,ambrosio2022ferroelectric,strelcov2017ch3nh3pbi3,li2022ferroelasticity,zheng2024ferroics}. Although ferroelastic domains are known to impact charge transport, recombination dynamics and carrier localization, their influence on excitonic structure remains poorly understood. In particular, it remains unclear whether ferroelastic symmetry breaking merely broadens optical resonances through disorder and strain or instead reorganizes the excitonic manifold itself by introducing symmetry-selective coupling pathways.

Addressing this problem experimentally is challenging because ferroelastic textures in conventional polycrystalline films are spatially disordered and optically isotropic\cite{strelcov2017ch3nh3pbi3,shrivastava2020polaron}, causing polarization-dependent excitonic responses to average out. Highly aligned anisotropic nanostructures circumvent this limitation by coupling ferroelastic texture to a well-defined optical axis, thereby enabling direct access to symmetry-selective excited-state dynamics through polarization-resolved spectroscopy.

\begin{figure*}
\centering
\includegraphics[width=\linewidth]{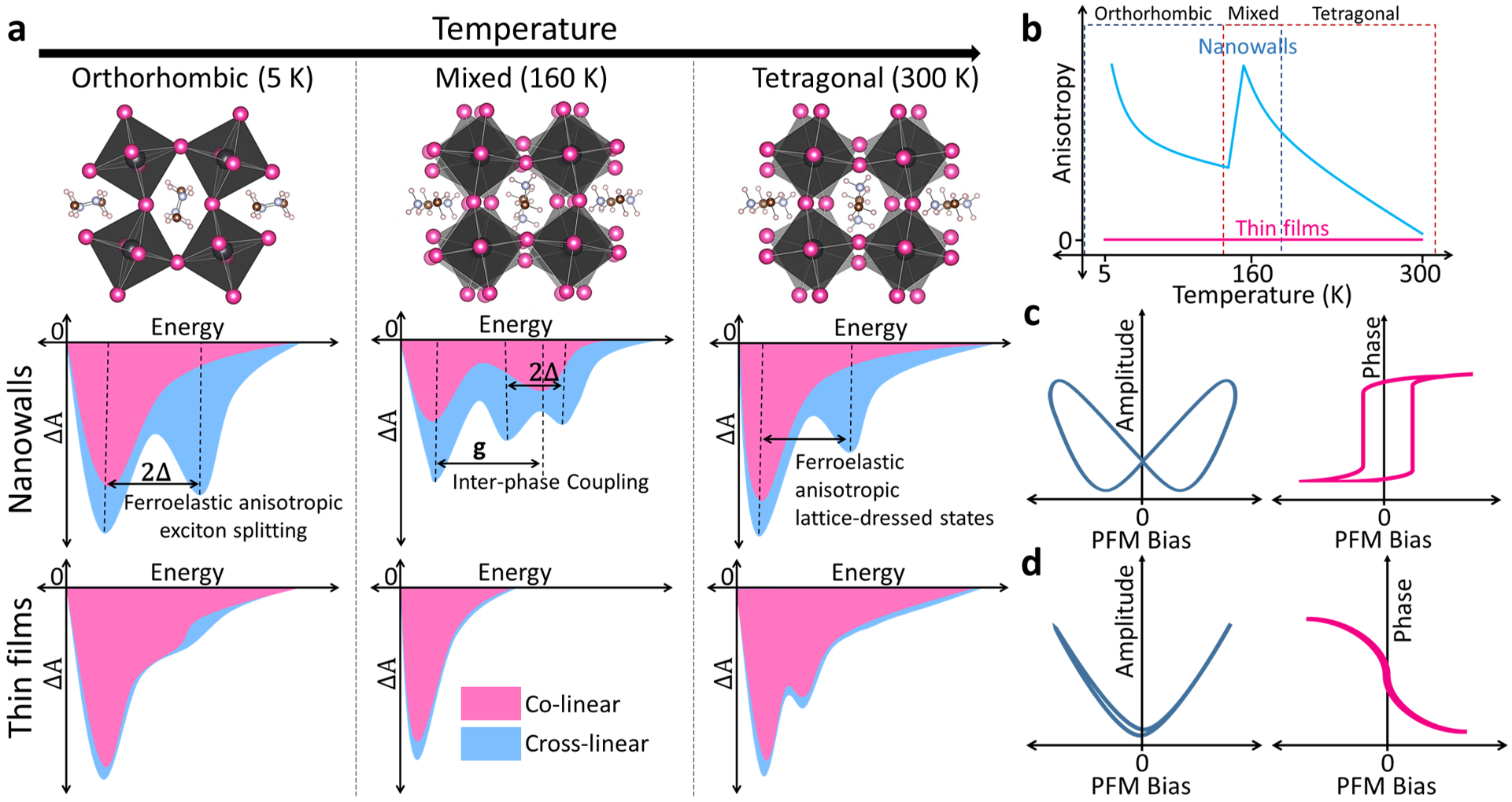}
\caption{1. Ferroelastic control of excitonic structure in highly aligned anisotropic MAPbI$_3$ nanowalls. (a) Schematic of the temperature-dependent structural phases and excitonic response in MAPbI$_3$ nanowalls (top) and planar thin films (bottom). At 5 K, the nanowalls adopt the orthorhombic phase, which hosts a polarization-selective excitonic splitting (2$\Delta$), revealed by polarization-resolved TA under co-linear (pink) and cross-linear (blue) pump-probe excitation. In the mixed orthorhombic-tetragonal regime  ($\sim$ 160 K), ferroelastic phase coexistence activates inter-phase coupling ($g$), producing a coupled excitonic manifold. At 300 K, the tetragonal phase is dominated by strongly lattice-dressed excitations accompanied by spectral broadening and loss of discrete excitonic character. By contrast, planar thin films show no polarization-selective splitting, highlighting the role of ferroelastic domain alignment in enabling symmetry-selective excitonic responses. (b) Temperature-dependent optical anisotropy. Nanowalls (blue) exhibit a distinct anomaly near the structural phase transition, whereas planar thin films (pink) remain largely isotropic. (c,d) PFM measurements of nanowalls (c) and thin films (d). Nanowalls exhibit clear amplitude hysteresis and phase switching under applied bias, consistent with ferroelastic-like behaviour, whereas planar thin films show no measurable switching response.}
\label{fig:1}
\noindent\rule{\textwidth}{0.4pt}
\end{figure*}

Here, we demonstrate symmetry-controlled evolution of excitonic and lattice-dressed states in highly aligned MAPbI$_3$ nanowalls fabricated by glancing-angle deposition\cite{hawkeye2014glancing,ai2018glancing}. The aligned nanowalls geometry generates a symmetry-selective ferroelastic environment in which polarization-resolved ultrafast transient absorption (TA) spectroscopy uncovers anisotropically split excitonic structure. Temperature-dependent photoluminescence, X-ray diffraction and TA measurements reveal an evolution from anisotropically split orthorhombic excitons to coupled excitonic and lattice-dressed states in the mixed-phase regime and finally to strongly lattice-coupled excitations in the tetragonal phase. In the orthorhombic phase, polarization-resolved TA reveals an excitonic splitting of 45 meV with orthogonal selection rules, while structural phase coexistence gives rise to a lower-energy lattice-dressed excitation 58 meV below the orthorhombic exciton. These results identify ferroelastic texture as a mechanism for tuning anisotropic exciton-lattice coupling and symmetry-selective excited-state response in soft polar semiconductors.

\begin{figure*}
\centering
\includegraphics[width=\linewidth]{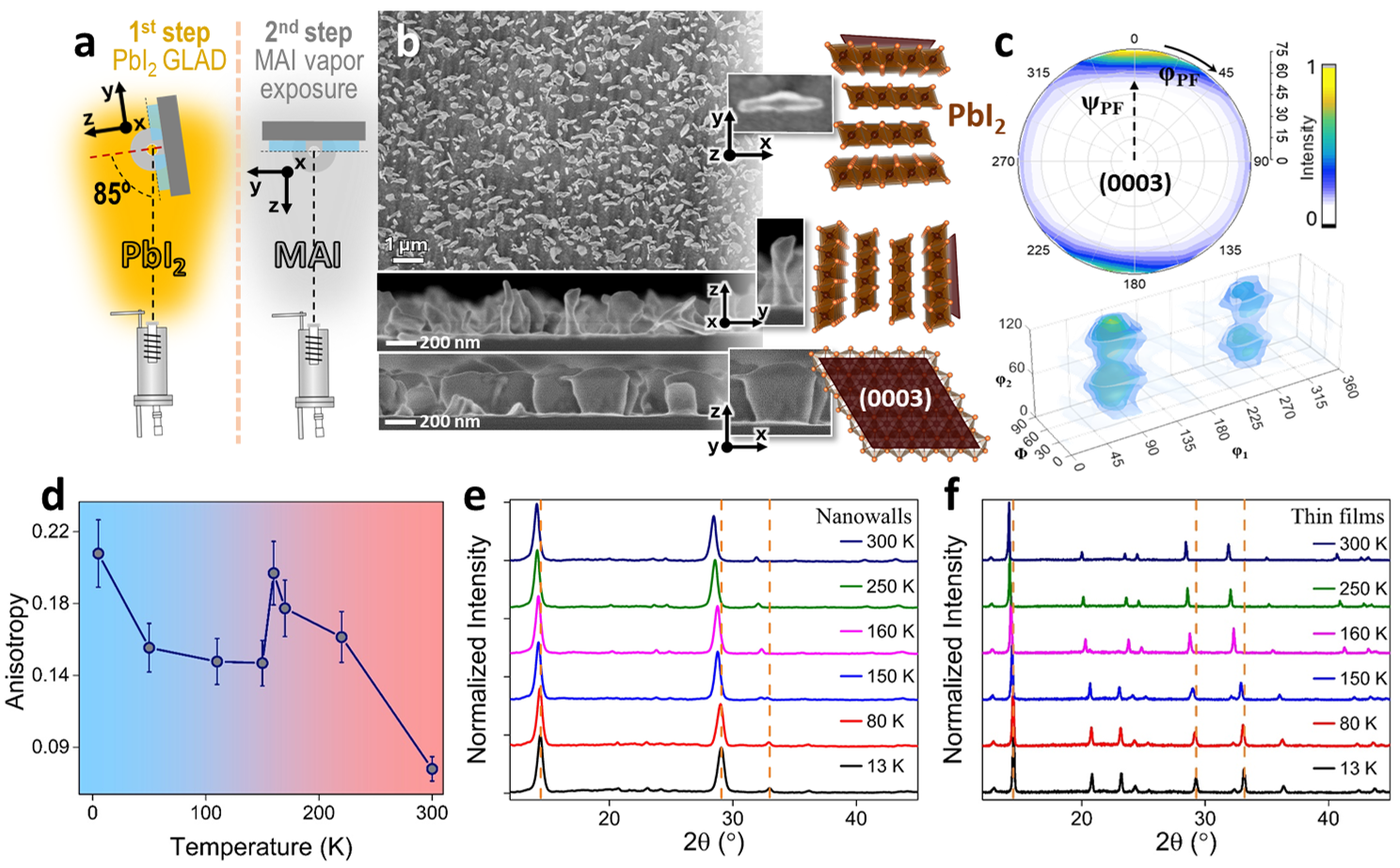}
\caption{2. Preparation and structural characterization of highly aligned MAPbI$_3$ nanowalls. (a) GLAD setup for anisotropic MAPbI$_3$ nanowalls: PbI$_2$ is deposited at an incident angle of 85$^\circ$, followed by MAI deposition at normal incidence. (b) Top-view (x-y) and cross-sectional (y-z and x-z) SEM images of MAPbI$_3$ nanowalls. (c) Experimental pole figure of the PbI$_2$ (0003) reflection (top) and the corresponding ODF reconstructed from the complete pole-figure set (bottom) (see also Supplementary Note 2). (d) Temperature-dependent optical anisotropy of nanowalls measured from 5-300 K, showing a distinct anomaly near 160 K, deviating from the otherwise monotonic temperature dependence. (e,f) Temperature-dependent XRD patterns of nanowalls (e), and planar thin films (f).}
\label{fig:2}
\noindent\rule{\textwidth}{0.4pt}
\end{figure*}

The core concept of our approach is illustrated in Fig. \ref{fig:1}a. Highly aligned MAPbI$_3$ nanowalls provide a symmetry-selective platform in which the temperature dependent orthorhombic to tetragonal structural phase transition reshapes the excited-state landscape through anisotropic exciton-lattice coupling. In the orthorhombic phase, polarization-resolved TA reveals anisotropically split excitonic states associated with ferroelastic symmetry breaking. Near the structural transition, coexistence of orthorhombic and tetragonal domains generates a mixed-phase regime characterized by enhanced exciton-lattice coupling and emergent low-energy excitations. At higher temperatures, these states evolve into strongly lattice-dressed excitations in the tetragonal phase. The aligned ferroelastic texture of the nanowalls introduces preferential optical axes and anisotropic strain environments that enable polarization-selective coupling pathways absent in planar thin films, producing strong temperature-dependent optical anisotropy (Fig. \ref{fig:1}b). Piezoresponse force microscopy further confirms the anisotropic ferroelastic character of the aligned nanowalls (Fig. \ref{fig:1}c) relative to planar thin films (Fig. \ref{fig:1}d). Together, these results identify ferroelastic texture and structural phase coexistence as mechanisms for tuning exciton-lattice coupling and symmetry-selective excited-state responses in soft polar semiconductors.\\
\\
\textbf{Symmetry-selective ferroelastic nanowalls architecture}\\
Motivated by the need to engineer ferroelastic anisotropy and symmetry-selective internal fields, highly aligned MAPbI$_3$ nanowalls film was fabricated using the glancing-angle deposition (GLAD) protocol\cite{hawkeye2014glancing,ai2018glancing} illustrated in Fig. \ref{fig:2}a (see Methods and Supplementary Note 1). The resulting architecture consists of vertically aligned nanowalls\cite{castillo2022highly} with strong in-plane anisotropy, whereas planar MAPbI$_3$ reference films were prepared by spin coating. Cross-sectional and top-view scanning electron microscopy (SEM) images (Fig. \ref{fig:2}b) reveal a dense array of highly aligned nanowalls preferentially oriented along the x-direction. Statistical analysis confirms well-separated walls with strong directional anisotropy (Supplementary Fig. 1). 

Beyond their anisotropic morphology, the nanowalls exhibit a strongly oriented crystallographic texture. Pole-figure analysis of PbI$_2$ nanowalls reveals preferential alignment of the crystallographic c-axis along a single in-plane azimuth with a tilt close to the surface normal (Fig. \ref{fig:2}c, Supplementary  Note 2 and Supplementary Fig. 2). Orientation distribution function (ODF)\cite{hielscher2008novel,bachmann2010texture} analysis further reveals a fibre-like texture with two symmetric tilt variants and minimal azimuthal dispersion, demonstrating that GLAD not only defines the macroscopic nanowalls morphology but also establishes a well-defined crystallographic orientation during growth, as schematically illustrated in Fig. \ref{fig:2}c.

To preserve structural integrity during ex-situ measurements, the nanowalls were encapsulated with a thin PMMA overlayer (Supplementary Fig. 3). Pole-figure and ODF analyses confirm that encapsulation induces only a geometrical tilting of the PbI$_2$ nanowalls while preserving the intrinsic GLAD-imposed crystallographic texture (Supplementary Note 2 and Supplementary Fig. 2)

The combined morphological and crystallographic anisotropy generates polarization-dependent optical anisotropy across the temperature range from 5 to 300 K (Supplementary Note 3 and Supplementary Fig. 4). The optical anisotropy parameter\cite{niu2018giant}, defined as (A$_{\parallel}-$A$_{\perp}$)/(A$_{\parallel}+$A$_{\perp}$), exhibits an anomaly near 160 K (Fig. \ref{fig:2}d), coincident with the orthorhombic to tetragonal structural phase transition. This anomaly is consistent with ferroelastic domain reorganization\cite{wadhawan2000introduction} and the emergence of temperature-dependent anisotropic ferroelastic environments within the aligned nanowalls ensemble.

To identify the structural origin of this anomaly, we performed temperature-dependent X-ray diffraction (XRD) measurements from 13 to 300 K. The diffraction patterns (Fig. \ref{fig:2}e and \ref{fig:2}f) reveal a strongly textured crystalline orientation in the nanowalls, in contrast to the polycrystalline character of planar thin films. Rietveld refinement\cite{sakata1979analysis,mccusker1999rietveld} further confirms the orthorhombic (\textit{Pnma})\cite{poglitsch1987dynamic} to tetragonal (\textit{I}4\textit{/mcm})\cite{poglitsch1987dynamic,whitfield2016structures} structural phase transition\cite{keshavarz2019tracking,li2023observation}. Between 120 and 180 K, both phases coexist in nanowalls and planar thin films, defining a mixed-phase regime that coincides with the temperature range of maximum optical anisotropy in the nanowalls. This directly links the optical anisotropy anomaly to ferroelastic phase coexistence.

The first-order tetragonal to orthorhombic phase transition in MAPbI$_3$\cite{whitfield2016structures} introduces coexistence of structurally distinct domains over a finite temperature window, generating significant lattice strain within the nanowalls ensemble. Williamson-Hall analysis\cite{williamson1953x} of diffraction peak broadening reveals a substantial increase in microstrain near the transition temperature (Supplementary Fig. 5), consistent with ferroelastic domain boundaries and associated lattice distortions (Supplementary Note 4). These strain fields are expected to produce anisotropic local lattice environments that couple to electronic excitations\cite{zou2024hysteresis}. Because the aligned nanowalls geometry imposes a preferred ferroelastic orientation, the resulting lattice distortions acquire a directional character, forming an anisotropic ferroelastic environment. This symmetry-selective coupling between lattice distortion and electronic excitations provides a physically plausible microscopic pathway through which ferroelastic symmetry breaking influences excitonic structure and optical anisotropy. 

\begin{figure}[t]
\includegraphics[width=\linewidth]{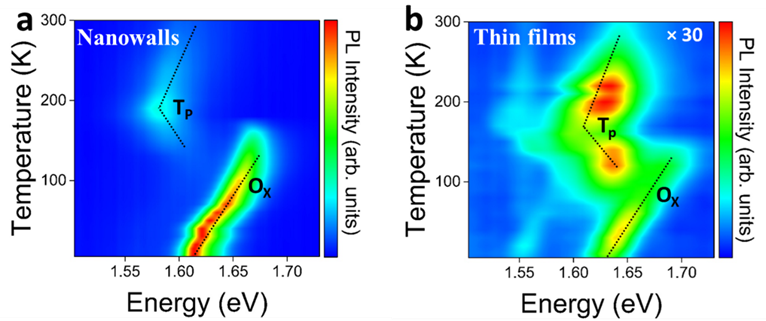}
\caption{3. Temperature-dependent evolution of excitonic and lattice-dressed states. (a) Temperature-dependent PL contour map of the MAPbI$_3$ nanowalls, showing the evolution from the low-temperature orthorhombic excitonic regime through the mixed-phase region to the high-temperature strongly lattice-dressed tetragonal regime. (b) Corresponding temperature-dependent PL contour map of the planar thin films. Although the overall spectral evolution is similar, the emission is substantially weaker and is therefore scaled by a factor of 30 for visibility.}
\label{fig:3}
\noindent\rule{\columnwidth}{0.4pt}
\end{figure}

Piezoresponse force microscopy (PFM) measurements further reveal preferentially aligned ferroelastic domain structures in the nanowalls\cite{leonhard2021evolution,garten2019existence}, in sharp contrast to the largely isotropic domain texture observed in planar thin films (Supplementary Note 5, Supplementary Figs. 6 and 7), consistent with an anisotropic ferroelastic texture imposed by the aligned nanowalls geometry.\\
\\
\textbf{Evolution of excitonic and lattice-dressed states}\\
To establish the equilibrium evolution of electronic excitations across the structural transition, we performed temperature-dependent photoluminescence (PL) measurements from 5 to 300 K. Figure \ref{fig:3}a shows that the nanowalls exhibit a sharp excitonic emission (O$_X$) at 5 K, with a full width at half maximum ($fwhm$ of 22 meV) consistent with Elliott analysis\cite{shrivastava2022room} (Supplementary Note 6 and Supplementary Fig. 8). Notably, O$_X$ exhibits an anomalous blueshift with increasing temperature, deviating from the conventional Varshni-type\cite{varshni1967temperature} behaviour expected for semiconductors\cite{bardeen1950deformation,saxena2020contrasting} and indicating strong exciton-lattice coupling in MAPbI$_3$ nanowalls. 

In the mixed-phase regime (120-180 K), an additional lower-energy band (T$_P$) emerges 50 meV below O$_{X}$, resulting in a distinct spectral splitting linked to phase coexistence. As the temperature increases further, O$_{X}$ weakens significantly while T$_P$ broadens and becomes dominant, consistent with increasing lattice dressing and enhanced carrier-phonon coupling in the tetragonal phase (see Supplementary Note 7 for detailed discussion). Rather than assigning T$_P$ to a uniquely defined microscopic polaron branch, we phenomenologically interpret it as a lattice-dressed excitation arising from strong exciton-lattice interactions within the mixed-phase and tetragonal regimes. Interestingly, the PL measurements in cooling and heating cycles exhibit hysteresis with memory effect characteristic of ferroelastic systems, consistent with hysteretic strain reconfiguration and ferroelastic domain evolution in MAPbI$_3$ nanowalls, see Supplementary Note 7 and Supplementary Fig. 9. In contrast, MAPbI$_3$ planar thin films exhibit similar spectral trends but with reduced intensity and diminished splitting, reflecting weaker anisotropic exciton-lattice interactions and reduced ferroelastic texture (Fig. \ref{fig:3}b, note the $\times$ 30 scale factor).

Density functional theory calculations combined with Monte Carlo sampling (Supplementary Note 8) qualitatively reproduce the experimentally observed trends in optical absorption and PL. First, the optical bandgap increases with temperature in both the orthorhombic and tetragonal phases (Supplementary Fig. 10), consistent with the PL measurements (Fig. \ref{fig:3}a). Second, the calculated optical anisotropy decreases from 0.16 in the orthorhombic phase to 0.08 in the tetragonal phase. Surface calculations for both MAI- and PbI$_2$-terminated surfaces further reveal local polar surface distortions within the top 2-3 layers\cite{Mohanty2026}. Whereas the MAI-terminated surface retains a nearly bulk-like anisotropy of $\sim$ 0.14, the PbI$_2$-terminated surface, shows an enhanced anisotropy of $\sim$ 0.24 with a smaller surface bandgap\cite{Mohanty2026}, consistent with enhanced surface polar distortions. This enhancement is experimentally observed in optical anisotropy (Fig. \ref{fig:2}d).\\
\\
\textbf{Ferroelastic symmetry breaking enables polarization-selective excitonic structure}\\
Having established the equilibrium excitonic landscape from PL measurements, we next investigate how ferroelastic symmetry breaking modifies excitonic states under nonequilibrium conditions using polarization-resolved ultrafast TA. Temperature-dependent TA measurements were performed using 3.1 eV pump excitation at a fluence of 250 $\mu$J/cm$^2$ (see Methods and Supplementary Note 9). By varying the pump-probe polarization between co-linear and cross-linear configurations, we decipher anisotropic optical responses arising from the aligned nanowalls geometry from polarization-independent excitonic contributions.

\begin{figure}[t]
\includegraphics[width=\linewidth]{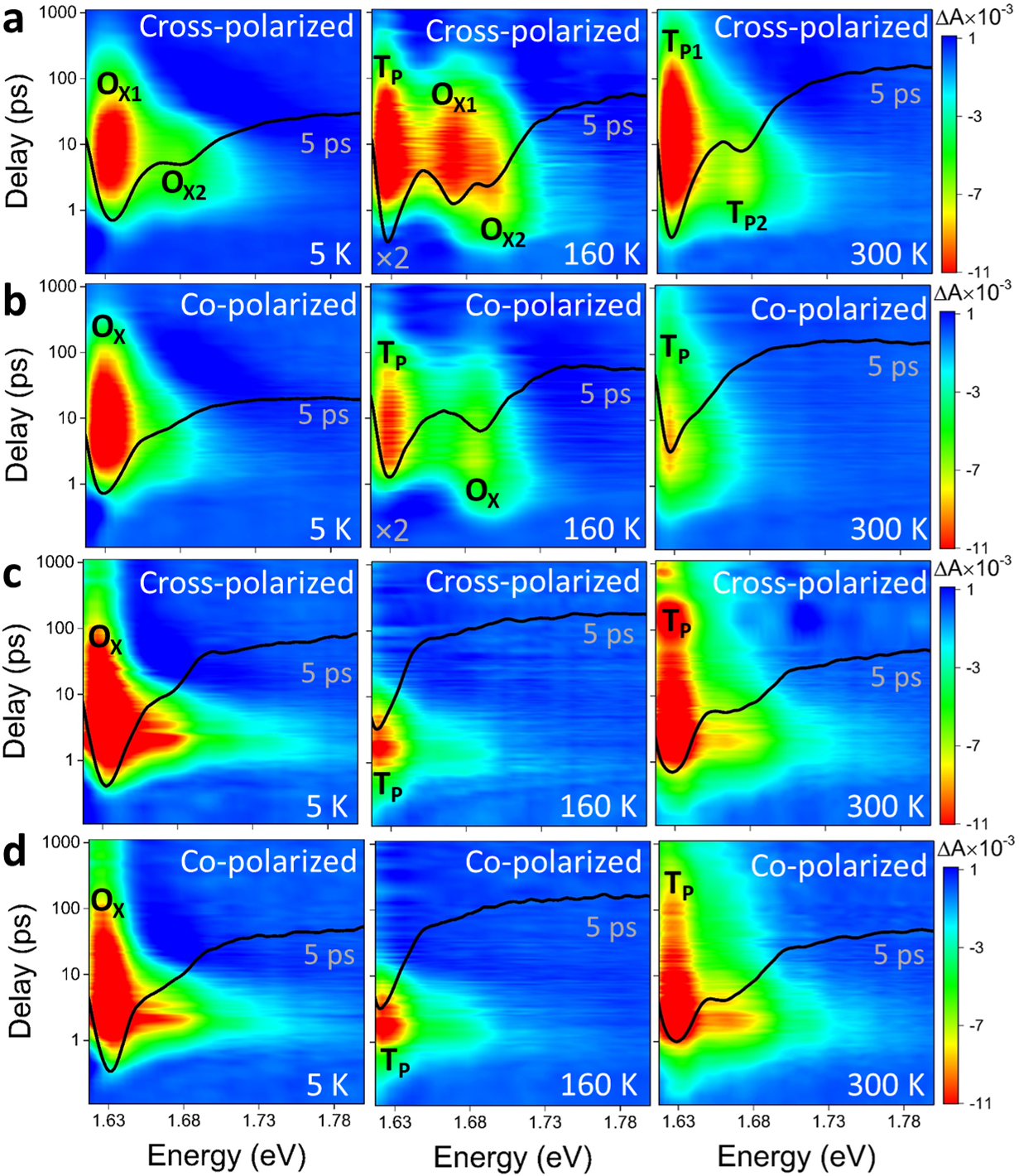}
\caption{4. Polarization-resolved TA reveals ferroelastic symmetry-controlled excitonic structure. (a) Cross-linear polarized TA contour maps of MAPbI$_3$ nanowalls at 5, 160, and 300 K, revealing pronounced excitonic splitting in the orthorhombic and mixed-phase regimes. At 300 K, the lattice-dressed response further exhibits internal spectral splitting. (b) Corresponding co-linear polarized TA contour maps of nanowalls measured under identical conditions, showing the absence of anisotropically split excitons and internal splitting of the lattice-dressed response. (c,d) Cross-linear (c) and co-linear (d) polarized TA contour maps of planar thin films under identical excitation conditions. Unlike the nanowalls, planar thin films do not exhibit polarization-dependent spectral splitting, highlighting the role of ferroelastic domain alignment and structural anisotropy in generating symmetry-selective excitonic responses. Black solid lines denote TA spectra extracted at a pump-probe delay of 5 ps.}
\label{fig:4}
\noindent\rule{\columnwidth}{0.4pt}
\end{figure}

Figure \ref{fig:4}a presents temperature-energy contour maps of the cross-linear TA spectra for the nanowalls, revealing the evolution of excitonic states across the structural transition. At 5 K, the spectra exhibit a well-resolved anisotropically split excitonic states with resonances at O$_{X1}$ at 1.631 eV and O$_{X2}$ at 1.676 eV. The energy separation between these states is 45 meV, significantly exceeding the excitonic \textit{fwhm} thereby ruling out thermal broadening or phonon-assisted effects as its origin. Instead, the splitting reflects a symmetry-broken anisotropic exciton response enabled by the aligned nanowalls geometry, which imposes a well-defined optical axis, allowing direct resolution of polarization-selective optical transitions accessible only in the cross-linear configuration. The absence of fluence-dependent energy shifts in O$_{X1}$ and O$_{X2}$  indicates that the anisotropic splitting is intrinsic rather than arising from many-body or pump-induced spectral effects.

Upon increasing temperature into the mixed-phase regime (160 K), the spectra evolve into a multicomponent structure indicating the coexistence of anisotropically split excitons in the orthorhombic phase and a low energy tetragonal lattice-dressed excitation (T$_P$). In this regime, the appearance of the T$_P$ introduces an additional channel, leading to coupling between orthorhombic excitons and lattice-dressed states. As a result, the apparent splitting between O$_{X1}$ and O$_{X2}$ reduces to 21 meV, accompanied by a redistribution of spectral weight and a partial loss of polarization contrast.

At 300 K, the spectra evolve into a broad lattice-dressed band that further exhibits internal splitting into T$_{P1}$ and T$_{P2}$, separated by 44 meV, indicative of strong carrier-phonon coupling in the tetragonal phase. Global kinetic analysis is provided in Supplementary Note 10 and Supplementary Fig. 11a. The extracted dynamics at 160 K show that O$_{X1}$ and O$_{X2}$ exhibit comparable lifetimes, attributed to their origin within the same excitonic manifold. By contrast, T$_P$ exhibits a longer lifetime than the anisotropically split excitons, supporting its assignment as a lattice-dressed excitation associated with tetragonal domains. At 300 K, T$_{P1}$ and T$_{P2}$ exhibit distinct relaxation dynamics, with the lower-energy T$_{P1}$ displaying a longer lifetime than T$_{P2}$, suggesting multiple lattice-dressed states with different relaxation channels.

In contrast to the cross-linear geometry, the co-linear TA spectra shown in Fig. \ref{fig:4}b exhibit a single excitonic bleach (O$_X$) at 1.626 eV at 5 K, similar to the PL response. Near 160 K, the spectra evolve into distinct O$_X$ and T$_P$ features separated by 62 meV, consistent with the mean exciton ((O$_{X1}$+O$_{X2}$)/2) -- T$_P$ separation observed in the cross-linear configuration. The anisotropic excitonic splitting observed in the cross-linear geometry is absent in the co-linear configuration, confirming its polarization-selective origin. At higher temperatures, O$_X$ is progressively suppressed while T$_P$ becomes dominant. Notably, the T$_{P1}$ -- T$_{P2}$ splitting observed at 300 K in the cross-linear configuration is absent in the co-linear geometry, indicating polarization-dependent selection rules for lattice-dressed states. The lifetime of T$_P$ in the co-linear configuration closely matches that of T$_{P1}$ in the cross-linear geometry (Supplementary Fig. 11b), supporting their correspondence.
 
 By contrast, planar thin films (Figs. \ref{fig:4}c and \ref{fig:4}d) do not exhibit polarization-dependent spectral splitting. These results demonstrate that ferroelastic symmetry breaking enables polarization-selective excitonic and lattice-dressed responses in MAPbI$_3$ nanowalls.\\
\\
\textbf{Mechanism}\\
To describe the temperature-dependent evolution of the excitonic structure, we employ a symmetry-guided effective 3$\times$3 Hamiltonian that incorporates anisotropically split excitons and interphase coupling between orthorhombic and tetragonal domains. The excitonic manifold is described in the basis of $\{ \ket{O_x}, \ket{O_y}, \ket{T} \}$ where $\{ \ket{O_x}$ and $\ket{O_y}$ denote the orthorhombic excitonic basis states and $\ket{T}$ represents the tetragonal lattice-dressed state,
\begin{eqnarray}
H(T) =
\begin{pmatrix}
E_O(T) + \Delta(T) & \delta_0(T) & g(T) \\
\delta_0(T) & E_O(T) - \Delta(T) & g(T) \\
g(T) & g(T) & E_T(T)
\end{pmatrix}
\label{eq:one}.
\end{eqnarray}

\begin{figure}[b]
\includegraphics[width=\linewidth]{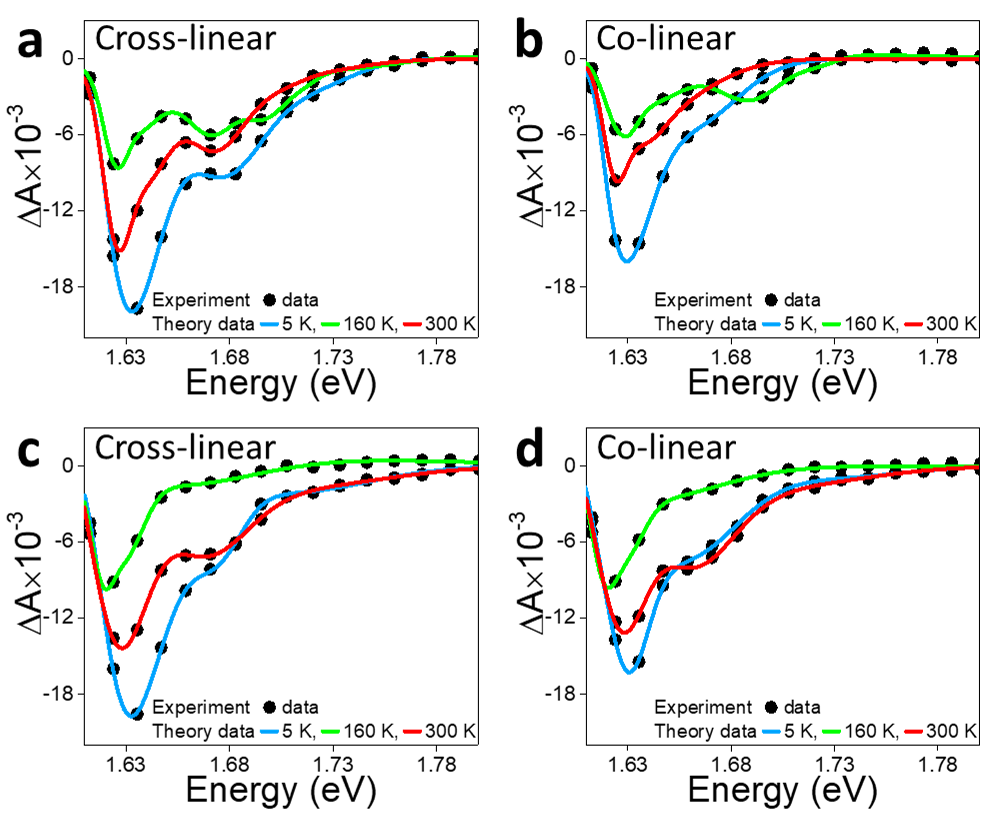}
\caption{5. Minimal coupled-exciton model and TA analysis. (a,b) Theoretical and experimental temperature-dependent TA spectra at a pump-probe delay of 5 ps for MAPbI$_3$ nanowalls. (c,d) Corresponding theoretical and experimental spectra for planar thin films. The theoretical calculations are based on the Hamiltonian defined in equation (\ref{eq:one}). The model reproduces the temperature-dependent evolution of the excitonic and lattice-dressed spectral features.}
\label{fig:5}
\noindent\rule{\columnwidth}{0.4pt}
\end{figure}
here $E_O(T)$ and $E_T(T)$ denote the mean energies of the orthorhombic excitonic manifold and tetragonal lattice-dressed state, respectively; 2$\Delta(T)$ describes the anisotropic exciton splitting; $\delta_0(T)$ is the intra-doublet mixing term within the orthorhombic manifold; and $g(T)$ parameterizes coupling between orthorhombic excitons and the tetragonal lattice-dressed state. Detailed estimation of the Hamiltonian parameters is provided in Supplementary Note 11. Fitting this model to the TA shown in Fig. \ref{fig:5} and Supplementary Fig. 12 captures the evolution from a well-resolved orthorhombic anisotropic exciton splitting at low temperature to a coupled manifold in the mixed-phase regime, followed by a crossover to predominantly lattice-dressed spectral weight at high temperature.\\
\\
\textbf{Discussion}\\
The temperature-dependent transient response of highly aligned MAPbI$_3$ nanowalls reveals that the excited-state manifold undergoes a progressive evolution across the orthorhombic-tetragonal transition, rather than simple thermal broadening. At 5 K, where the tetragonal contribution is negligible, polarization-resolved TA reveals anisotropically split excitonic states separated by 45 meV with orthogonal optical selection rules. The splitting substantially exceeds the low-temperature \textit{fwhm} and is absent in planar thin films, supporting its origin in the symmetry-selective ferroelastic texture imposed by the aligned nanowalls geometry, consistent with previous studies of ferroelastic twins in hybrid perovskites\cite{ambrosio2022ferroelectric,whitfield2016structures,li2023observation}. In this regime, the response is well described by a reduced 2$\times$2 Hamiltonian corresponding to anisotropically split orthorhombic excitons. The pump-probe polarization angle ($\phi$) dependent change in oscillator strength of $\delta_O$ shown in Supplementary Fig. 13, support the effective Hamiltonian. 

This picture changes in the mixed-phase regime between 120 and 180 K, where orthorhombic and tetragonal domains coexist. Here, an additional lower-energy excitation emerges 58 meV below the centre of the orthorhombic excitonic manifold, accompanied by spectral weight redistribution and reduced polarization contrast. Within the effective Hamiltonian framework, these observations are consistent with coupling between anisotropic orthorhombic excitons and lattice-dressed excitations associated with tetragonal domains. The coexistence regime therefore provides an environment in which structural anisotropy, ferroelastic texture and exciton-lattice interactions become strongly intertwined.

At higher temperatures, the discrete orthorhombic excitons evolve into broadened lattice-dressed excitations characteristic of the tetragonal phase, accompanied by increased phonon occupation and dynamic disorder. The optical response is more naturally described in terms of strongly lattice-dressed excitations than well-defined excitons.  

The observed excitonic restructuring is not readily explained by conventional strain or electron-phonon coupling alone. Strain-induced effects typically produce continuous energy shifts or modest spectral splitting, whereas electron-phonon interactions predominantly lead to spectral broadening and phonon sidebands. By contrast, the MAPbI$_3$ nanowalls exhibit discrete polarization-selective excitonic states with large energy separation and well-defined optical selection rules that emerge only in the presence of aligned ferroelastic domains and are absent in planar thin films measured under comparable conditions. Furthermore, the appearance of an additional excitation branch in the mixed-phase regime is consistent with the coupling between orthorhombic excitons and tetragonal lattice-dressed states, rather than simple phonon dressing.

In summary, the interplay of ferroelastic domains, anisotropic strain, and phase coexistence in MAPbI$_3$ nanowalls drives a temperature-dependent evolution of excitonic and lattice-dressed states. This reorganization proceeds from anisotropically split excitons at low temperature to a strongly coupled excitonic manifold near the phase boundary, before crossing over to predominantly lattice-dressed excitations in the tetragonal phase. A symmetry-guided effective Hamiltonian captures the essential features of this evolution. These results establish aligned hybrid perovskite nanowalls as a platform for exploring symmetry-selective exciton-lattice coupling in soft semiconductors.\\
\\
\textbf{References}\\
\bibliographystyle{naturemag}
\bibliography{MAPI}
\vspace{4mm}
\textbf{Methods}\\
\\
\textbf{Sample preparation}\\
PbI$_2$ and MAI were thermally evaporated from separate thermal evaporation sources to form highly aligned MAPbI$_3$ nanowalls. PbI$_2$ was deposited at $\sim$ 330$^\circ$ C under < 1 $\times$ 10$^{-5}$ mbar onto fused silica and Si(100) substrates at an 85$^\circ$ glancing angle (15 cm distance), with a growth rate of $\sim$ 0.1 nm s$^{-1}$ monitored by quartz crystal microbalance. MAI was subsequently sublimated at $\sim$ 160$^\circ$ C (18 cm distance) with the substrate normal to the flux under $\sim$ 1 $\times$ 10$^{-4}$ mbar, achieved via partial throttling of the turbopump. Complete conversion to MAPbI$_3$ nanowalls was obtained after 60 min. Samples were transferred without air exposure and encapsulated with a 5 wt $\%$ PMMA layer (spin-coated at 2000 rpm for 10 s and 5000 rpm for 30 s, $\sim$ 150 $\mu$L), followed by annealing at 70$^\circ$ C for 30 min.\\ 
Reference MAPbI$_3$ films were prepared using a DMF/DMSO precursor with 1.3 M concentrations of PbI$_2$ and MAI. PbI$_2$ was dissolved at 70$^\circ$ C, followed by MAI addition at room temperature. Films were deposited by spin coating at 5000 rpm for 50 s, with diethyl ether antisolvent applied after 6 s, and annealed at 65$^\circ$ C for 1 min and 100$^\circ$ C for 2 min to complete crystallization.\\
\\
\textbf{Characterization Methods}\\
High-resolution scanning electron microscopy (SEM) images were recorded with a Hitachi S4800 system operated at an accelerating voltage of 2 kV. Cross-sectional micrographs were obtained after mechanically cleaving the silicon substrates.\\
The pole figures were recorded using a Malvern Panalytical Empyrean diffractometer equipped with a 5-axis Eulerian cradle, allowing independent control of the tilt angle ($\psi_{\mathrm{PF}}$) and the azimuthal rotation ($\phi_{\mathrm{PF}}$) of the sample. Cu K$\alpha$ radiation ($\lambda = 1.5418$~\AA{}) was used. The diffraction angle ($2\theta$) was fixed at the positions corresponding to the (0003), (11$\bar{2}$0), and (10$\bar{1}$0) reflections of PbI$_2$ nanowalls. The $\psi_{\mathrm{PF}}$ angle was varied from $0^\circ$ to $75^\circ$, while $\phi_{\mathrm{PF}}$ was scanned over the full $0^\circ$--$360^\circ$ range. Data were collected using a step size of $5^\circ$ in both $\psi_{\mathrm{PF}}$ and $\phi_{\mathrm{PF}}$. Before analysis, standard corrections for background and defocusing effects were applied to the measured intensities. Data processing, including background and defocusing corrections, as well as the calculation of the orientation distribution functions (ODFs), was performed using the open-source MTEX software package.\\
The XRD patterns were obtained with a Panalytical X’Pert Pro X-ray diffractometer with a Cu K$\alpha$ radiation source with cryostat ranging from 13-300 K.\\
Piezo-response force microscopy (PFM) measurements were performed at RT under ambient conditions using a Park NX-10 atomic force microscope (Park Systems). A conductive Pt/Ir-coated probe ($k = 3$ N/m, $f_0 \sim 75$ kHz) was used. PFM imaging was carried out in contact mode by applying an AC voltage to the tip to induce local piezoelectric vibrations. The out-of-plane (vertical) piezoresponse was detected using an internal lock-in amplifier. PFM writing-reading experiments were conducted to probe spatially resolved switching. A reference image was first acquired using only the AC signal. Subsequently, alternating polarization patterns were written by applying local DC biases of $\pm$ 20 V and $\pm$ 25 V through the tip, with the AC excitation disabled. After removing the DC bias, the same area was rescanned under AC excitation to confirm stable domain patterns.\\
Temperature-dependent ultrafast transient absorption (TA) spectroscopy was performed using a Ti:sapphire oscillator (Mai Tai), generating $\sim$ 100 fs pulses centered at 800 nm. The oscillator output was used to seed a regenerative amplifier, producing amplified pulses with an average energy of $\sim$ 4 mJ at a repetition rate of 1 kHz with a pulse width of $\sim$ 120 fs. The high-intensity beam was directed into a $\beta$-barium borate (BBO) crystal for second-harmonic generation to produce the 400 nm pump beam. Then the beams were separated using a dichroic beam splitter. The 800 nm beam was passed through a computer-controlled delay stage and subsequently focused into a calcium fluoride (CaF$_2$) crystal to generate a white-light continuum spanning 440-775 nm. The pump and probe beams were spatially overlapped on the sample, and the transient change in absorbance was measured. For co-linearly and cross-linearly polarization configuration, we used a half wave plate to change the excitation polarization. All measurement done under identical excitation conditions with a pump fluence of 250 $\mu$J/cm$^2$.\\
\\
\textbf{Acknowledgment}\\
K.V.A. gratefully acknowledges the Anusandhan National Research Foundation (project no. CRG/2023/003013) and the Department of Science and Technology Nano Mission (project no. DST/NM/TUE/QM-8/2019(G/1)). Afreen acknowledges the Ministry of Education for support from the Prime Minister's Research Fellowship (PMRF ID: 0403017). J.R.S.V. and the Nanotechnology on Surfaces \& Plasma Laboratory acknowledge funding from projects PID2022-143120OB-I00 and PCI2024-153451 (ANGSTROM), funded by MCIN/AEI/10.13039/501100011033 and by ``ERDF (FEDER), A way of making Europe'', ``Fondos NextGenerationEU'', and ``Plan de Recuperaci\'on, Transformaci\'on y Resiliencia''. Project ANGSTROM was selected in the Joint Transnational Call 2023 of M-ERA.NET 3, an EU-funded network of about 49 funding organizations (Horizon 2020 grant agreement No.~958174). Additional support was provided by project DGP\_PIDI\_2024\_02239 funded by the European Union, ``Ministerio de Hacienda'', European Funds and ``Junta de Andaluc\'ia'', Intramural CSIC Project 202460E235, and the EU Horizon 2020 program under grant agreement No.~851929 (ERC Starting Grant 3DScavengers).
\\
\\
\textbf{Author contribution}\\
K.V.A. conceived the research and supervised the experiments. Afreen performed optical absorption, PL and TA measurements. P.T. helped in polarization dependent optical and transient absorption measurements. J.D.A., J.C.S. synthesized the MAPbI$_3$ nanowalls and thin films. J.R.S.V, A.Bo, and A.Ba characterize the layers by SEM and XRD pole figures and interpret the data. H.K.M performed the PFM characterization. DFT calculations was performed by P.M. and K.M. All authors contributed to writing, discussing and editing the paper.\\
\\
\textbf{Competing interests}\\
The authors declare no competing interests.

\end{document}